\documentstyle[12pt]{article}
\advance\textheight by \topskip
\begin{document}

\rightline{SU-ITP-93-35}
\rightline{MSUHEP-93-23}
\rightline{UCSD/PTH-93-45}
\rightline{Dec 1993}

\vspace{.8cm}
\begin{center}
{\large\bf  Next-to-leading Order Parton Model Calculations
in the Massless Schwinger Model\\}

\vskip .9 cm

{\bf Jin Dai
\footnote{E-mail address: dai@higgs.ucsd.edu} }
 \vskip 0.1cm
Department of Physics 0319,
University of California, San Diego \\
9500 Gilman Dr.
La Jolla, CA, 92093-0319

{\bf James  Hughes
\footnote{E-mail address: hughes@msupa.pa.msu.edu} }
 \vskip 0.1cm
Physics and Astronomy Department,
Michigan State University, \\
East Lansing, MI  48823

{\bf
Jun Liu
\footnote{E-mail address: junliu@scs.slac.stanford.edu} }
 \vskip 0.1cm
Physics Department,
Stanford University,
Stanford, CA 94305

\end{center}

\vskip .6 cm
\centerline{\bf ABSTRACT}
\vspace{-0.7cm}
\begin{quotation}
We carry out next-to-leading order (NLO)
parton model calculations for the standard hard ``QCD''
processes in the massless Schwinger model.
The asymptotic expansion of the
exact result for the deep inelastic cross section is
used to infer the NLO distribution function.
These distribution functions are then used to calculate
the NLO Drell-Yan parton model cross section and
it is found to agree with the corresponding
term in the expansion of the exact result for the
Drell-Yan cross section.
Finally, by using the bosonization formula and the
exact solutions we study the interference
between different partonic processes.

\end{quotation}

\normalsize
\newpage

\section{Introduction }

Although the
parton model\cite{bj,feynman} lacks a solid theoretical foundation,
it is an essential tool in QCD applications to strong interactions
and so far it has been reasonably
successful. If we have a theory that exhibits many properties of
4-dimensional quantum chromodynamics, ${\rm QCD}_4$, and
this theory is exactly solvable, then it is important that we
study how well parton model works in this theory.  The massless Schwinger
model is one such theory in that it is
an exactly solvable, interacting quantum
field theory \cite{sch,lowenstein}
that is both asymptotically free and exhibits
confinement
\cite{coleman,suss}.

In reference \cite{hugliu1}, the exact cross sections for lepton-antilepton
annihilation, deep inelastic scattering and Drell-Yan processes in
the massless Schwinger model coupled to a scalar current are
calculated in terms of the functions $R_{\pm}(q^2)$.
In \cite{hugliu2} the full asymptotic expansions of
$R_\pm (q^2)$ were described and the terms up to and including
order $(g^2/q^2)^4$ were explicitly computed where $g$ is the
strong coupling constant and $q^2$ is the squared
momentum transfer. The leading terms
in these exact cross sections were then shown to equal the leading
order parton model results in the Bjorken scaling region.
This paper extends this analysis of the massless Schwinger model
to the next-to-leading order (NLO).

Of course, the lack of transverse
momenta for the Schwinger model partons
and the corresponding absence of jets
limits the extent of the lessons that we adduce for ${\rm QCD}_4$.
Another fundamental difference between
the Schwinger model and ${\rm QCD}_4$ is
that the Schwinger model coupling constant $g$ has the dimensions of
mass.  However, we see this difference as an opportunity to
consider several potential problems for the parton model
when it is is pushed to next order
in $g^2/q^2$ including: (1) the mass of the hadron which
is of order $g$ in this model can no longer be neglected in the
kinematics; (2) hadronization or bound state effects are assumed
to be suppressed by inverse powers of $q^2$ and so may arise at
order $g^2/q^2$; (3) quantum interference of the hard partonic
processes are normally assumed to be power suppressed and so may
arise at order $g^2/q^2$; and finally (4) the analogue
of higher twist terms.

To our surprise, despite these complications, we find that the parton
model NLO Drell-Yan cross section agrees with the exact Drell-Yan
cross section at NLO. We also argue that at NLO the Schwinger model photon,
the analogue in our model of the gluon, has a parton distribution,
which may be calculated by evaluating processes at order $g^4/q^4$.
As another application, we use
the bosonization formula to isolate the contributions
coming from the interference of the underlying partonic
processes. Then from the expansion of the exact
results, we find that the interference is suppressed
by order $(g^2/q^2)^4$.

\section{NLO Parton Distribution Functions From Deep Inelastic Scattering}

The processes that we calculate occur in the same model used
in \cite{hugliu1,hugliu2,coleman,suss}.
For completeness we briefly describe
the model here.  We extend the massless Schwinger model
\cite{sch} by including a
massless fermion $f$ (our ``lepton'') that is not ${\rm QED}_2$-charged
but interacts with the ${\rm QED}_2$-charged fermions, $\psi$, through
a Yukawa coupling to a scalar photon $\phi$.  The full Lagrangian is
\cite{coleman,suss}
\begin{eqnarray}
L={F^2}/4+\bar{\psi}(i\not\!{\partial}+g\not\!\!{A})\psi\label{lagran}
+\bar{f}i\not\!{\partial}f+1/2\phi\Box\phi
+e(\bar{\psi}\psi+\bar{f}f)\phi.
\end{eqnarray}
So $e$ is the analogue of the electromagnetic coupling,
and $g$ is the analogue of the ${\rm QCD}_4$ strong coupling constant
so from now on,we will call the photon of QED2 as ``gluon''. All of
our calculations are done at lowest order in $e$.  Then, using the
dual realization of the Schwinger model in terms of a free
scalar of mass( which we will call ``hadron'') $m_h=g/\sqrt {\pi}$
and the bosonization formula
\footnote{The prefactor $c$ is a normal ordering dependent constant
and it equals $c=\frac{g \gamma }{(2\pi )^3/2}$ when the
normal ordering mass equals $m_h$.}
\begin{eqnarray}
\bar{\psi}\psi=c:\cos(2\sqrt{\pi}h):, \label{boson}
\end{eqnarray}
we may compute the process cross sections exactly in $g$.

First consider the DIS process
$$
f(k_1)+h(P)\rightarrow f(k_2)+X,
$$
where $h$ is the target particle and the final states $X$ are summed
over and where the momentum assignments are given in the
parantheses.  The cross section at leading order in $e^2$, but all orders in
$g^2$, is
\begin{eqnarray}
d\sigma
=\frac{e^4}{q^4}\frac{1}{2E_P}R_5((q+P)^2)dk^\prime.
\end{eqnarray}
where $q=k_1-k_2$ is the momentum transfer, and $P$ is the hadron
momentum, $R_5(q^2)$ is defined by
\begin{eqnarray}
R_5(q^2)=(R_{+}(q^2)-R_{-}(q^2))/2
\end{eqnarray}
and
\begin{eqnarray}
R_{\pm}(q^2)=c^2\int d^2x \exp(iqx) \exp (\pm 4\pi\Delta_m(x)),\label{rpm}
\end{eqnarray}
In \cite{hugliu2} we showed that for $q^2\neq 0$,
\begin{eqnarray}
R_{+}(q^2)=1+\frac {1}{2\pi ^2}(g^2/q^2)^2+(g^2/q^2)^3
\frac{1}{\pi ^3}(12+4\ln [\pi g^2/q^2])+o((g^2/q^2)^4),
\label{rplus}
\end{eqnarray}
and
\begin{eqnarray}
R_{-}(q^2)=(g^2/q^2)^4
\frac{1}{\pi ^4}(6.96+3.79\ln [\pi g^2/q^2])+o((g^2/q^2)^5).
\label{rmin}
\end{eqnarray}
Thus the cross section in DIS has no $o(g^2/q^2)$ correction to the
leading order result. We will use this and the parton
cross sections presented next to evaluate the NLO parton
distribution functions.

The parton model calculation of DIS is done perturbatively in
$g$  where the final states that are summed over consist of
quanta of $\psi$ and of $A^{\mu}$.  The vector field $A^{\mu}$ has
asymptotic propagating states because we have chosen to
regulate the IR collinear divergences (the same fermion mass
singularities that occur in perturbative ${\rm QCD}_4$ calculations) by
temporarily giving the $A^{\mu}$ field a mass, $m_g$.
At leading order in $g$ the only process that contributes
is
$$
f(k_1)+\psi(p_1)\rightarrow f(k_2)+\psi(p_2),
$$
and the cross section is given in \cite {hugliu1} to be
\footnote{Throughout this paper, we denote the cross-sections of
the partonic subprocess, the parton model calculation,
and the exact calculation by  $\hat{\sigma}$, $\bar{\sigma}$, and $\sigma$
respectively.}
\begin{eqnarray}
  d\hat{\sigma}^0=\frac{e^4}{4q^4}\frac{dk_2}{E_1}\delta(1-z).
          \label{dips}
\end{eqnarray}
where $z \equiv -q^2/(2p_1 \cdot q)$.

As in ${\rm QCD}_4$, the NLO $g^2$ corrections come from
the interference of the one loop corrections with the leading order
process, $d\hat{\sigma}_v^1$,
and from gluon bremstrahlung, $d\hat{\sigma}_b^1$.
We choose to regularize by giving the ``gluon'' a mass $m_g$.
The calculational steps that go into evaluating
$d\hat{\sigma}_v^1$ are almost the same as in those outlined in
\cite {hugliu2}.  The result, written in terms of
$\beta \equiv {m_g^2}/{q^2} $, is found to be
\begin{eqnarray}
  d\hat{\sigma}_v^1=\frac{e^4}{4q^4}\frac{dk_2}{E_p}\delta(1-z)
  \frac{g^2}{q^2} \frac{1}{\pi \beta} \ln [-\beta] .\label{dipsv}
\end{eqnarray}

The calculation of the bremstrahlung\footnote
{2-dimension gauge bosons have no physical degree of freedom, however,
after regulariztion, the Schwinger model phonton obtains a longetudinal
mode. This makes the model much more like QCD.
This method has proven successful in the calculation of
lepton-antilepton annihilation process. (\cite{hugliu2}) }
cross section is a little more involved and warrants a brief description.
The process is
$$
f(k_1)+\psi(p_1)\rightarrow f(k_2)+\psi(p_2)+A^{\mu}(p_3),
$$
and the corresponding cross section is directly calculated to be
\begin{eqnarray}
d\hat{\sigma}_b^1=\frac{e^4}{4q^4}\frac{dk_2}{E_p}
\frac{g^2}{q^2} \frac{1}{2\pi}\frac{(-8q^4)}{(p_1-p_3)^2(p_2-p_3)^2}
p_1\cdot p_2
\frac{dp_2}{E_2}
\frac{dp_3}{E_3}
\delta^2(k_1+p_1-k_2-p_2-p_3)
.\label{dipsb}
\end{eqnarray}
It is an exercise in kinematics to do the integrals over $p_2$, $p_3$
and rewrite this cross section in terms
of the two independent variables $q^2$ and $z$.
The result is found to be
\begin{eqnarray}
d\hat{\sigma}_b^1=\frac{e^4}{4q^4}\frac{dk_2}{E_p}
\frac{g^2}{q^2} \frac{1}{\pi \beta}\frac{1}{1-z}
.\label{dipsbz}
\end{eqnarray}
For this radiation correction it is important to note that the condition
\begin{eqnarray}
(p_1+q)^2\ge m_g^2,
\end{eqnarray}
implies
\begin{eqnarray}
\frac{1}{1-\beta}\ge z. \label{lim}
\end{eqnarray}

Now we are in a position to fold the partonic cross sections into
a hadronic result. First notice that the mass of the hadron is of order
$g$, so we cannot neglect this mass in the next-to-leading order
calculation. We will follow the well-established method to deal with
the hadron mass\cite{xisc}. We define, without loss of generality,
the hadron momentum fraction $y$ carried by the parton as
\begin{eqnarray}
p_{1+} = yP_+,
\end{eqnarray}
where
\[p_1^2=0 \; , \;\; q_+=q_0 + q_1 \; , \;\; q_-=q_0 - q_1  \; .\]
It follows that $z=\xi / y$, where $\xi$ is the Nachtmann scaling variable
\begin{eqnarray}
  \xi \equiv  \frac{2x}{1 + \sqrt{1 - 4x^2m_h^2/q^2}}
\end{eqnarray}
Also the familiar Bjorken scaling variable $x$ is defined to be
\begin{eqnarray}
    x\equiv -\frac{q^2}{2P \cdot q},
\end{eqnarray}

The hadronic cross section in the parton model is
then given by multiplying the underlying partonic cross sections by
the parton distribution functions, both
given as a functions of $\xi$, $y$, and $q^2$,
and integrating over $y$ over its allowed
limits.  Thus
\begin{eqnarray}
d\bar{\sigma}=\sum_i\int_0^1 dy f_i(y,q^2)d\hat{\sigma}(y,q^2).
\label{partonb}
\end{eqnarray}

The partonic cross sections in equations (\ref {dips}),(\ref {dipsv}), and
(\ref {dipsbz}) must be summed to give $d\hat{\sigma}$ to the
desired order.  Further charge conjugation invariance of the
underlying theory implies that the fermion and antifermion
distributions are equal.  Hence the sum over $i$ in (\ref{partonb})
is replaced by a factor of $2$.  Finally, the leading order
result $f=1$ is all that is needed in evaluating the integral
over $y$ with the NLO cross sections
$d\hat{\sigma}_v^1$ and
$d\hat{\sigma}_b^1$.
The only subtlety in the integration over $y$
occurs in the integration limits for
the bremstrahlung cross section.  There
equation (\ref{lim}) implies that
\begin{eqnarray}
{(1-\beta )}\xi \le y\le 1. \label{blim}
\end{eqnarray}
Setting the partonic result equal
to the exact result and keeping terms up to order $g^2/q^2$ gives
\begin{eqnarray}
f(y,q^2)=1-\frac{g^2}{q^2} \frac{1}{\pi \beta} \ln [\frac{1-y}{y}].
\label{NLOf}
\end{eqnarray}
This is the full NLO distribution function for
Schwinger model partons in the DIS process.  Recall that in
${\rm QCD}_4$ the NLO distribution functions
may be characterized in terms of the Altarelli-Parisi equations \cite{AP}.
These differential equations describe the
running of the NLO distribution functions with
$q^2$ due to the collinear divergence of the radiative
gluon corrections.  The initial conditions for the distribution
functions
must come from a comparison with experimental data at some
fixed $q_0^2$.  In the calculations of this section, we have
effectively solved for both the running and the initial conditions
by directly relating the cross sections.

Observe that the momentum sum rule is {\it not} satisfied by these
NLO distributions.  Indeed,
\begin{eqnarray}
2\int_0^1 dy yf(y,q^2)=1+\frac{g^2}{m_g^2}\times (-\infty). \label{mom}
\end{eqnarray}
{}From this we conclude that, in the massive ${\rm QED}_2$-gluon
regularization scheme, the
gluon has a distribution whose expansion in $g$ starts at
order $g^2/q^2$.  To calculate this distribution from DIS would
require going to order $g^4/q^4$ since the contributing cross section
is itself of order $g^2/q^2$.  Instead, we will now use the NLO
distribution functions inferred from DIS to evaluate
the Drell-Yan cross section.

\section{Drell-Yan Process at Next-to-leading Order}

We next use the  parton model to
calculate the cross-section  for the Drell-Yan process,
$$
h(P)+h(P^\prime)\rightarrow f(k)+\bar{f}(k^\prime )+X.
$$
In the parton model it is
assumed that the distribution functions $f(x,q^2)$ used in the
calculation of DIS are the same
as the ones $f^{DY}(x,q^2)$ to be used in the calculation of Drell-Yan process
\cite{georgi}.
But, a prior, these  distribution functions $f^{DY}(x,q^2)$
with timelike momentum transfer, $q^2>0$,
and $f(x,q^2)$ for the
DIS process with spacelike momentum tranfer, $q^ 2<0$, could be
different.
In \cite{hugliu1} it is shown that they are both equal to 1.  However,
higher order radiation might spoil this equality.
In fact, the exchange of soft gluons between the initial state hadrons
is often cited as a possible complication in even applying
the parton model to the Drell-Yan process.
Nevertheless, we will verify
in this section.
that the NLO parton model
cross section, with the initial
state radiation included, equals the exact cross section at NLO.

The cross section to leading order in $e^2$ but exact in $g^2$ is
given in \cite {hugliu1} to be
\begin{equation}
d\sigma=\frac{e^4}{q^4}
        \frac{1}{ \left[(P \cdot P^\prime)^2 - g^4/{\pi^2}\right]
        ^{\frac{1}{2}} } dkdk^\prime R((q-P-P^\prime )^2),
\end{equation}
where
\begin{eqnarray}
R(q^2)&=&R_+(q^2)+R_-(q^2) \nonumber \\
&=&c^2\int d^2x \exp(iqx) \cosh (4\pi\Delta_m(x)).\label{r}
\end{eqnarray}
Then using the results of the asymptotic expansion in equations
(\ref{rplus},\ref{rmin}) we see that the exact
Drell-Yan cross section is
\begin{eqnarray}
d\sigma=\frac{e^4}{q^4}
\frac{dkdk^\prime}{P \cdot P^\prime}
(1+{\rm O}(\frac{g^4}{q^4})).
\label{xsecexact}
\end{eqnarray}

To evaluate the NLO parton model, Drell-Yan cross section we must
evaluate the underlying partonic cross sections and then fold
them together with the parton distributions in the incident hadrons.
At leading order the parton process is
$$
\psi (p)+\bar {\psi}(p^\prime)\rightarrow f(k)+\bar{f}(k^\prime ),
$$
and the cross section is evaluated in \cite {hugliu1} to be
\begin{eqnarray}
d\hat{\sigma}^0_{DY}=
\frac{e^4}{2q^4}
dkdk'
\delta^2(p+p^\prime-k-k^\prime).
\label{DY0}
\end{eqnarray}
To go to NLO we need to
evaluate the radiative corrections to the leading order
cross section.
As with DIS, there are two corrections.  One from
the interference of the one-loop amplitudes with the leading
order process, $d\hat {\sigma}^1_{DYv}$.
And the other from the emmission of a final state
gluon from one of the incoming partons,
$d\hat {\sigma}^1_{DYb}$.

The calculation of the virtual cross section follows the same
lines as the calculations of the virtual corrections to
DIS - given in the preceding section -
and of the virtual corrections to $f-\bar {f}$ annihilation described
in \cite {hugliu2}.  The result is that
\begin{eqnarray}
d\hat{\sigma}^1_{DY,v}=
d\hat{\sigma}^0_{DY}
\frac {g^2}{q^2}\frac{1}{\pi \beta}\ln |\beta|.
\label{DY1v}
\end{eqnarray}

The other order $g^2/q^2$ correction comes from the process
$$
\psi (p)+\bar {\psi}(p^\prime)\rightarrow f(k)+\bar{f}(k^\prime )
+A^{\mu}(r),
$$
and the corresponding cross section is immediately calculated
to be
\begin{eqnarray}
d\hat{\sigma}^1_{DY,b}=
\frac{e^4}{2q^4}
dkdk'
\delta^2(k+k^\prime+r-p-p^\prime)
\frac{dr}{E_r}
\frac {g^2}{q^2}\frac{1}{\pi}\frac{q^4}{(p'-r)^2(p-r)^2}.
\label{DY1b}
\end{eqnarray}
These partonic cross sections are then folded together with
the NLO distribution functions to get the
NLO parton model cross-section as
follows
\begin{eqnarray}
d\bar{\sigma}&=&\sum_{i,j}\int_0^1 dydy^\prime f_i(y)f_j(y^\prime)
d\hat{\sigma}_{i,j}(y,y^\prime)
dkdk^\prime
\nonumber\\
&=&
2\int_0^1 dydy^\prime f(y)f(y^\prime)
[d\hat{\sigma}^0_{DY}
+d\hat{\sigma}^1_{DY,v}
+\int dr \frac{d\hat{\sigma}^1_{DY,b}}{dr}].
\label{pDY}
\end{eqnarray}
In the second line of equality, we have used charge conjugation
invariance to replace the sum over parton species with a factor of 2.
Also, the outgoing gluon momentum must be integrated over in
the bremstrahlung contribution. We will evaluate the contributions
from the three cross sections separately.

For the first two terms:
\begin{eqnarray}
 & &2 \int_0^1 dydy^\prime  f(y)f(y^\prime) \left( d\hat{\sigma}^0_{DY}
            +  d\hat{\sigma}^1_{DY,v} \right)   \nonumber \\
 &=& \frac{e^4}{q^4} dkdk^\prime \int_0^1 dydy^\prime
       \delta^2(q - p - p^\prime)
        \left\{ 1 -\frac{g^2}{q^2}\frac{1}{\pi\beta}
        \left[ \ln\frac{1-y}{y} + \ln\frac{1-y^\prime}{y^\prime} \right]
        + \frac{g^2}{q^2}\frac{1}{\pi\beta} \ln\beta \right\} \;\;\;
\end{eqnarray}
where
\[   p_+ = yP_+ \; , \;\;
   p^\prime_- = y^\prime P^\prime_-  \; . \]
The integration over $y$ and $y^\prime$ can be easily done using the
following identity:
\begin{equation}
   \delta^2(q) = 2\delta(q_+)\delta(q_-)   \label{delta}
\end{equation}
and we get:
\begin{eqnarray}
  & &2 \int_0^1 dydy^\prime  f(y)f(y^\prime) \left( d\hat{\sigma}^0_{DY}
            +  d\hat{\sigma}^0_{DY,v} \right)   \nonumber \\
  &=& \frac{e^4}{q^4}\frac{1}{PP^\prime} dkdk^\prime
     \left[ 1 - \frac{g^2}{q^2}\frac{1}{\pi\beta} \ln\left(
     \frac{2PP^\prime - 2Pq -2P^\prime q + q^2}{q^2\beta} \right)\right]
  \label{dy0v}
\end{eqnarray}

Finally, consider the contribution due to the bremstrahlung cross
section, $d\hat{\sigma}^1_{DY,b}$.
We need only the leading order distribution function, and at this order,
we can neglect the effect of the mass of the hadron, but now the integral
is complicated by the extra integral over the gluon momemtum, $r^{\mu}$.
Using equation (\ref{delta}), we can do the integrals over $y$ and $y'$
first, and after some algebras, get:
\begin{equation}
    2 \int_0^1 dydy^\prime f(y)f(y^\prime)
     \int dr \frac{\hat{\sigma}^1_{DY,b}}{dr}
   = \frac{e^4}{q^4}\frac{1}{PP^\prime} dkdk^\prime
       \frac{g^2}{q^2}\frac{1}{\pi\beta} \int\frac{dr}{E_r}
\end{equation}
We are left with the integral over the momentum $r$.  The
limits of the integration are determined by the conditions that
$y<1$ and $y'<1$  which imply that
\begin{equation}
   -\frac{(P^\prime_- - q_-)^2 - m_g^2}{2(P^\prime_- - q_-)}
      \leq r \leq \frac{(P_+ - q_+)^2 - m_g^2}{2(P_+ - q_+)}
\end{equation}
Therefore, the contribution to the parton model Drell-Yan cross
section due to gluon emission is given by
\begin{equation}
   2 \int_0^1 dydy^\prime f(y)f(y^\prime)
       \int dr \frac{\hat{\sigma}^1_{DY,b}}{dr}
   = \frac{e^4}{q^4}\frac{1}{PP^\prime} dkdk^\prime
   \frac{g^2}{q^2}\frac{1}{\pi\beta} \ln\left(
     \frac{2PP^\prime - 2Pq -2P^\prime q + q^2}{q^2\beta} \right)
   \label{dyb}
\end{equation}

Comparing equations (\ref{dy0v}) and (\ref{dyb}), we find that the
infinity in the parton distribution function cancels with that in
Drell-Yan parton cross section, and the net
NLO correction to the Drell-Yan cross-section in the parton model
calculation vanishes:
\begin{equation}
  d\hat{\sigma}_{DY}=d{\sigma}_{DY}
       =\frac{e^4}{q^4}\frac{1}{PP^\prime} dkdk^\prime
\end{equation}
That is, the NLO parton model calculation of the Drell-Yan
cross section, using the NLO distribution functions inferred from
the DIS process, equals the exact Drell-Yan cross section evaluated
to next-to-leading order.

\section{Discussion and Conclusion}

Simple dimension counting in ${\rm QCD}_4$ distinguishes the
perturbative corrections according to their twist.
Typically, one focuses on the perturbative
large logarithmic corrections and
neglects the higher twist corrections that are suppressed
by powers of $1/q^2$.
In contrast, the solvable $1+1$-dimensional massless Schwinger model
coupling constant $g$ has the dimensions
of mass so that all of the corrections are suppressed by
powers of $g^2/q^2$.  Thus the perturbative corrections can not be
cleanly separated from corrections due to hadron mass, interference and
hadronization, as well as potential higher twist terms.
We have exploited this difference to study these effects in the
parton model.

It turns out the $\xi$ scaling handles the hadron mass correctly
in this model. However, it is kind of a surprise that we get a
Drell-Yan cross section which agrees precisely with the exact
solution without introducing multi-parton densities. There is
still much to understand in this model.

The calculations in this paper establish the universality
of the distribution functions
in the Schwinger model at next-to-leading order.  Using
the asymptotic expansion of the exact results we can test
whether this process-independence persists to even higher
orders.  Other assumptions of the model can also be tested.  For
example, in \cite{hugliu2} we argued that the $(g^2/q^2)^4\ln [g^2/q^2]$
term in the expansion of the exact annihilation cross section
was an effect of hadronization that can not be calculated from
perturbation theory (see equation (\ref{rmin}).
In ${\rm QCD}_4$ it is tacitly assumed that these hadronization
effects are supressed by inverse
powers of the large squared-momentum transfer.
Another fundamental assumption of the parton model is that the
interference of different
partonic processes is suppressed by inverse powers of large
squared-momentum transfer.  For example, this assumption is
built into the starting point of those proofs of
various factorization theorems that rely on Landau-Cutovsky
cut diagrams \cite{collins}.
At least in one case, we can test this assumption in the Schwinger model.

First, separate the density in equation (\ref{boson})
that couples to the scalar
photon into two parts:
\begin{equation}
  \bar\psi \psi = \bar\psi_L \psi_R + \bar\psi_R \psi_L .\label{lr}
\end{equation}
Let us use $a$ and $b$ to represent the annihilation operators for
the quarks and antiquarks, respectively. Then
\begin{eqnarray}
   \bar\psi_L \psi_R \sim a^\dagger_L a_R + b^\dagger_R b_L + \cdots
             \nonumber \\
   \bar\psi_R \psi_L \sim a^\dagger_R a_L + b^\dagger_L b_R + \cdots
            .
\end{eqnarray}
Next, without loss of generality, assume the hadron is moving right. Then
the contribution of the first term of equation (\ref{lr}) to DIS process
corresponds to a quark coming out of the hadron and the second term
corresponds to an antiquark coming out the hadron. In parton model,
the cross sections for these two process are summed over. In the exact
solution, however, we can calculate the interference of these two terms
using
\begin{eqnarray}
    \bar\psi_L \psi_R = \frac{c}{2}:exp(2i\sqrt{\pi}h)  \nonumber \\
    \bar\psi_R \psi_L = \frac{c}{2}:exp(-2i\sqrt{\pi}h)
\end{eqnarray}
Thus, it turns out the inference term is proportional to $R_-(q^2)$ and
from equation (\ref{rmin}) the leading order of this interference occurs
at
$o(g^8/q^8)$. This term is indeed is very small but is nonvanishing.
A similar analysis applies to Drell-Yan
process.



\section*{Acknowledgements}
We have benefited from discussions with S. Brodsky, J. Bjorken,
A. Manohar, M. Peskin, H. D. Politzer, L. Susskind, and C.P. Yuan.
J.H. was supported during part of this work
by the US DOE under Contract No. W-7405-ENG-48(LLNL) and the Nuclear
Theory Grant No. SF-ENG-48.
J.D. was supported by DOE under grant DE-FG03-90ER40546.
\vskip 1cm


\end{document}